\documentclass{IEEEtran}
\usepackage{cite}
\usepackage{amsmath,amssymb,amsfonts}
\usepackage{algorithmic}
\usepackage{graphicx}
\usepackage{textcomp}
\usepackage{algorithmic}
\usepackage{algorithm}
\usepackage{array}
\usepackage[caption=false,font=normalsize,labelfont=sf,textfont=sf]{subfig}
\usepackage{stfloats}
\usepackage{url}
\usepackage{verbatim}
\usepackage{enumerate}
\usepackage{tikz}
\usepackage{bm}
\usepackage{threeparttable,booktabs}
\def\BibTeX{{\rm B\kern-.05em{\sc i\kern-.025em b}\kern-.08em
    T\kern-.1667em\lower.7ex\hbox{E}\kern-.125emX}}
\begin{document}
\title{Synthesis Method for Obtaining Characteristic Modes of Multi-Structure Systems via independent Structure T-Matrix}

\author{Chenbo Shi, Xin Gu, Shichen Liang, Jin Pan and Le Zuo
\thanks{Manuscript received Sep. 24, 2024. This work was supported by the Aeronautical Science Fund under Grant ASFC-20220005080001. (\textit{Corresponding author: Jin Pan.})}
\thanks{Chenbo Shi, Xin Gu, Shichen Liang and Jin Pan are with the School of Electronic Science and Engineering, University of Electronic Science and Technology of China, Chengdu 611731 China  (e-mail: chenbo\_shi@163.com; xin\_gu04@163.com; lscstu001@163.com; panjin@uestc.edu.cn).}
\thanks{Le Zuo is with The 29th Research Institute of China Electronics Technology Group Corporation (e-mail: zorro1204@163.com)}
}

\markboth{ }%
{Shell \MakeLowercase{\textit{et al.}}: A Sample Article Using IEEEtran.cls for IEEE Journals}

\maketitle

\begin{abstract}
This paper presents a novel and efficient method for characteristic mode decomposition in multi-structure systems. By leveraging the translation and rotation matrices of vector spherical wavefunctions, our approach enables the synthesis of a composite system's characteristic modes using independently computed simulations of its constituent structures. The computationally intensive translation process is simplified by decomposing it into three streamlined sub-tasks: rotation, z-axis translation, and inverse rotation, collectively achieving significant improvements in computational efficiency. Furthermore, this method facilitates the exploration of structural orientation effects without incurring additional computational overhead. A series of illustrative numerical examples is provided to validate the accuracy of the proposed method and underscore its substantial advantages in both computational efficiency and practical applicability.
\end{abstract}


\section{Introduction}

\IEEEPARstart{T}{he} theory of characteristic modes is pivotal in antenna analysis and design \cite{ref_review1,ref_review2,ref_review3}. Its derived form---substructure characteristic modes theory---has gained prominence for revealing the intrinsic electromagnetic properties of structures within complex environments \cite{ref_sTCM_Concept,ref_sTCM_Huang1,ref_sTCM_Review}. This approach is increasingly employed in designing antennas for handheld devices and platform-mounted systems \cite{ref_sTCM_Huang2}. Recent extensions have unified various characteristic mode formulations, which were traditionally based on the method of moments (MoM), into a more robust field-domain framework, significantly enhancing numerical efficiency \cite{ref_CM_unify,ref_my}. Despite these advancements, the computational burdens for systems with multiple structures, such as antenna arrays, remain substantial.

In the field-domain framework \cite{ref_CM_unify,ref_my}, characteristic modes are obtained by eigenvalue decomposition of the transition matrix (T-matrix) or scattering matrix. This matrix serves as an operator that connects the incidence and structural scattering responses, solely determined by the properties of the structures within a designated enclosing sphere. This characteristic indicates the possibility to construct the T-matrix for combined structures from independent structural data, thus facilitating the rapid computation of characteristic modes for complex systems. This paper primarily focuses on optimizing and exploring this potential.

Although extensive research in fields such as optical scattering has effectively elucidated and summarized the technique of deriving the total system's T-matrix from the individual structures' T-matrices \cite{ref_Tmat_1,ref_Tmat_2,ref_Tmat_3,ref_Tmat_4}, particularly through employing the two translation properties of spherical wavefunctions \cite{ref_sph_addition1,ref_sph_addition2}, we revisit this established method from a unique perspective. We have developed a fully matrixized representation that is not only concise and efficient but also facilitates easier understanding compared to conventional series representations. This is particularly advantageous in scenarios that require specifying the radiation background for characteristic mode decomposition, where our approach minimizes repetitive computations. Following the method outlined in \cite{ref_sph_near_measure}, we decompose the general translation problem into three sub-steps: rotation, z-axis translation, and inverse rotation, eschewing the direct solutions typically employed in conventional studies. This significantly curtails computation time, as z-axis translations are inherently simpler. Moreover, by adjusting the translation direction using the rotation matrix, we can correlate problems that have identical translation distances but different directions---common in uniformly arranged structures---thereby enhancing the reuse of the translation matrix and further boosting computational efficiency.

The integration of the rotation matrix within our method introduces additional degrees of freedom; notably, it enables alterations to the structure's orientation during post-processing, obviating the need to recompute the T-matrix. This enhancement is crucial for investigating the effects of the structure's posture or polarization. To substantiate our approach, we present a series of illustrative numerical examples that demonstrate the substantial advantages of our synthesis technique in deriving characteristic modes. These examples highlight the potential applications of our method in complex structures and antenna arrays.

\section{Characteristic Modes Theory in Field Domain}
\label{SecII}

As a foundational pillar of this study, this section provides a concise overview of the latest advancements in characteristic mode theory, equipping readers with the theoretical framework employed throughout this paper.

The field-domain formulation of characteristic mode theory articulates its foundational equation using the scattering matrices of structures \cite{ref_my}:
\begin{equation}
  \label{eq1}
  \mathbf{SS}_{b}^{\dagger}\mathbf{f}_n=s_n \mathbf{f}_n \text{ or }\left(\mathbf{SS}_{b}^{\dagger}-\mathbf{1}\right)\mathbf{f}_n=2t_n \mathbf{f}_n.
\end{equation}
Given the relationships $\mathbf{S}=\mathbf{1}+2\mathbf{T}$ and $\mathbf{S}_b=\mathbf{1}+2\mathbf{T}_b$, \eqref{eq1} can be reformulated as
\begin{equation}
  \label{eq2}
  \left(\mathbf{T}+\mathbf{T}_{b}^{\dagger}+2\mathbf{T}\mathbf{T}_b^\dagger\right)\mathbf{f}_n=t_n \mathbf{f}_n.
\end{equation}
Here, $\mathbf{f}_n$ denotes the expansion coefficients of the 4-type (outgoing) vector spherical wavefunctions for the $n$-th characteristic radiation field, $\mathbf{T}_b$ represents the background's T-matrix, and $\mathbf{T}$ corresponds to the total system's T-matrix, encompassing both the key and background structures.

The presence of a background structure $\Omega_b$ allows \eqref{eq2} to solve for the characteristic mode of the key structure $\Omega_k$ within an environment inclusive of $\Omega_b$. This scenario is often referred to as the ``substructure characteristic mode''. The most basic form occurs when $\mathbf{T}_b=\mathbf{0}$, in which case \eqref{eq2} solves for the characteristic mode of the structure radiating in free space.

The algebraic connection between \eqref{eq2} and the MoM is derived from pre-assuming wavefunctions based on real spherical harmonics \{cf. \cite{ref_scattering_theory} for definitions of these wavefunctions\}: 
\begin{equation}
  \label{eq3}
  \mathbf{T}=-\mathbf{PZ}^{-1}\mathbf{P}^t\text{, }\mathbf{T}_b=-\mathbf{P}_b\mathbf{Z}_{b}^{-1}\mathbf{P}_{b}^{t}
\end{equation}
where $\mathbf{P}$ and $\mathbf{P}_b$ are matrices that relate the equivalent electromagnetic currents to their respective radiation fields. While the field-domain approach does not rely on traditional MoM equations, \eqref{eq3} remains pivotal for computing T-matrices. The distinctions among various CM formulations have been comprehensively explored in \cite{ref_CM_unify,ref_my}---with solutions derived from \eqref{eq2} generally demonstrating superior performance.

The maximum degree of spherical waves required for accurate representation is estimated using the truncation criterion \cite{ref_trunction}:
\begin{equation*}
  L_{\max }=\left\lceil k r_\text{max}+2  \sqrt[3]{k r_\text{max}}+3\right\rceil
\end{equation*}
where $r_\text{max}$ represents the farthest radial distance from the structure to the origin. This results in a total of $2L_{\max }(L_{\max }+2)$ spherical wavefunctions being employed.

\section{Synthesis of the T-Matrix for Multi-Structure Systems}
\label{Sec_III}

In applications involving multiple structures, calculating the T-matrix of the total system using \eqref{eq3} can be burdensome. However, it is possible to synthesize the system T-matrix using individual structure T-matrices, somewhat akin to domain decomposition methods. The T-matrix maps the 1-type spherical wavefunction coefficients $\mathbf{a}$ of incident waves to the 4-type coefficients $\mathbf{f}$ of radiating (scattering) waves, \textit{i.e.,} $\mathbf{f}=\mathbf{T}\mathbf{a}$. This allows for an understanding of the system's T-matrix in relation to its components through a scattering-problem perspective.

\begin{figure}[!t]
  \centering
  \includegraphics[]{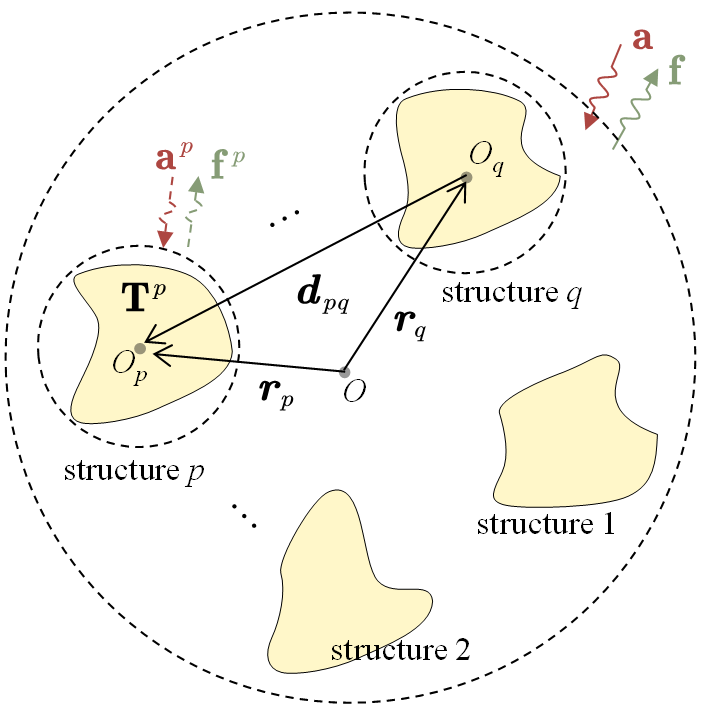}
  \caption{Schematic of electromagnetic scattering for a multi-structure system.}
  \label{fScatterModel}
\end{figure}

Assuming the system comprises $M$ structures, labeled as $p=1,2,\cdots,M$ (see Fig.~\ref{fScatterModel}), the system generates a scattering wave $\mathbf{f}$ under an external incident wave $\mathbf{a}$. From the perspective of the $p$-th structure (using its local coordinate system), the contribution to the scattering wave reads
\begin{equation}
  \label{eq4}
  \mathbf{f}^p=\mathbf{T}^p\mathbf{a}^p.
\end{equation}

The incident wave $\mathbf{a}^p$ for the $p$-th structure comprises a direct component from $\mathbf{a}$ (denoted by $\mathbf{a}^p_d$), and scattering contributions $\mathbf{f}^q$ from other structures ($q \ne p$). By translating $\mathbf{f}^q$ into the $p$-th local coordinate system, we express it as: 
\begin{equation}
  \label{eq5}
  \mathbf{a}^p=\mathbf{a}^p_d+\sum_{q \ne p}^M \bm{\mathcal{Y}}_{pq}^t \mathbf{f}^q
\end{equation}
where $\bm{\mathcal{Y}}_{pq}$ represents the translation matrix. Substituting this into \eqref{eq4}, we obtain
\begin{equation}
  \label{eq6}
  \mathbf{f}^p=\mathbf{T}^p\mathbf{a}^p_d+\mathbf{T}^p \sum_{q \ne p}^M \bm{\mathcal{Y}}_{pq}^t \mathbf{f}^q.
\end{equation}

If we further transform \eqref{eq6} into matrix form, we have
\begin{equation}
  \tilde{\mathbf{f}}=\tilde{\mathbf{T}}\tilde{\mathbf{a}}_d+\tilde{\mathbf{T}}\tilde{\bm{\mathcal{Y}}}\tilde{\mathbf{f}}.
\end{equation}
This leads to
\begin{equation}
  \label{eq8}
  \tilde{\mathbf{f}}=\left( \mathbf{1}-\tilde{\mathbf{T}}\tilde{\bm{\mathcal{Y}}} \right) ^{-1}\tilde{\mathbf{T}}\tilde{\mathbf{a}}_d
\end{equation}
where 
\begin{equation*}
  \tilde{\mathbf{f}}=\begin{bmatrix}
    \mathbf{f}^1\\
    \vdots\\
    \mathbf{f}^M\\
  \end{bmatrix} ,\tilde{\mathbf{a}}_d\begin{bmatrix}
    \mathbf{a}_{d}^{1}\\
    \vdots\\
    \mathbf{a}_{d}^{M}\\
  \end{bmatrix} ,
  \tilde{\bm{\mathcal{Y}}}=\left[ \begin{matrix}
    \mathbf{0}&		\bm{\mathcal{Y}} _{12}^{t}&		\cdots&		\bm{\mathcal{Y}} _{1M}^{t}\\
    \bm{\mathcal{Y}} _{21}^{t}&		\mathbf{0}&		\cdots&		\bm{\mathcal{Y}} _{2M}^{t}\\
    \vdots&		\cdots&		\ddots&		\vdots\\
    \bm{\mathcal{Y}} _{M1}^{t}&		\bm{\mathcal{Y}} _{M2}^{t}&		\cdots&		\mathbf{0}\\
  \end{matrix} \right] 
\end{equation*}
and $\tilde{\mathbf{T}}=\mathrm{diag}\left( \mathbf{T}^1,\mathbf{T}^2,\cdots ,\mathbf{T}^M \right)$.

For the moment we have successfully represented the radiated fields generated by each structure within their respective local coordinate systems. To determine the T-matrix for the total system, we must correlate these localized quantities into the global coordinate system. This necessitates the introduction of another translation matrix, $\bm{\mathcal{R}}_p$, leading to $\mathbf{a}_d^p=\bm{\mathcal{R}}_p^t \mathbf{a}$ and
\begin{equation*}
  \mathbf{f}=\sum_p^M{\bm{\mathcal{R}}_p \mathbf{f}^p}.
\end{equation*}
In matrix form, this can be simplified to
\begin{equation}
  \label{eq9}
  \mathbf{f}=\tilde{\bm{\mathcal{R}}}\tilde{\mathbf{f}}\text{, }\tilde{\mathbf{a}}_d=\tilde{\bm{\mathcal{R}}}^t \mathbf{a}
\end{equation}
where $\tilde{\bm{\mathcal{R}}}=\left[\bm{\mathcal{R}}_1,\bm{\mathcal{R}}_2,\cdots, \bm{\mathcal{R}}_M\right]$. 

By substituting \eqref{eq9} into \eqref{eq8} and comparing with the expression $\mathbf{f}=\mathbf{T}\mathbf{a}$, we can derive the system T-matrix:
\begin{equation}
  \label{eq10}
  \mathbf{T}=\tilde{\bm{\mathcal{R}}} \left( \mathbf{1}-\tilde{\mathbf{T}}\tilde{\bm{\mathcal{Y}}} \right) ^{-1}\tilde{\mathbf{T}} \tilde{\bm{\mathcal{R}}}^t.
\end{equation}
Note that details regarding the translation matrices $\bm{\mathcal{Y}}$ and $\bm{\mathcal{R}}$ will be discussed in the upcoming section.

Through carefully examining each matrix in \eqref{eq10}, we can see that the T-matrix of the total system is synthesized from the T-matrices of its constituent structures independently, along with translation matrices that relate solely to positional topology. Notably, this approach does not introduce any additional information about mutual couplings between different structures; these couplings are inherently captured by the spherical wavefunctions, which represents a significant advantage of this theory. In contrast to conventional methods for obtaining the T-matrix of multiple structures, our formulation \eqref{eq10} is fully matrixized, easier to implement, and computationally efficient (see Sec. \ref{SecV_A} for data support of this conclusion).

Another essential task is to determine the T-matrix of the background, $\mathbf{T}_b$. This is synthesized using the T-matrices of the remaining structures after excluding the key structures. The process is similar to the one described above, with the primary difference being the removal of matrix blocks associated with the key structures. If we prioritize computing the background $\mathbf{T}_b$, the total system T-matrix can then be obtained via the Schur complement method (cf. appendix \ref{app_A}), which eliminates much of the redundant computation.

\section{Translation and Rotation Matrix for Spherical Wavefunctions}

In this section, we explore the methods for computing the translation matrices $\bm{\mathcal{Y}}_{pq}$ and $\bm{\mathcal{R}}_p$. Both matrices are functions related to the electrical translation distance, \textit{i.e.}, $\bm{\mathcal{Y}}_{pq}=\bm{\mathcal{Y}}(k \boldsymbol{d}_{pq})$, $\bm{\mathcal{R}}_{p}=\bm{\mathcal{R}}(k \boldsymbol{r}_{p})$, with $\bm{\mathcal{R}}=\mathrm{Re}\left\{\bm{\mathcal{Y}}\right\}$. Here, $\bm{\mathcal{Y}}$ matrix translates 4-type spherical wavefunctions to 1-type, whereas $\bm{\mathcal{R}}$ handles translations among the same type of spherical wavefunctions. This underpins the use of $\bm{\mathcal{Y}}_{pq}$ in \eqref{eq5}---to translate the scattering wave (4-type) into the incident wave (1-type).

Representations for $\bm{\mathcal{Y}}(k \boldsymbol{d})$ can be found in existing literature. A notable simplification occurs when $\boldsymbol{d}$ is parallel to the z-axis, where $\bm{\mathcal{Y}}(k \boldsymbol{d})$ becomes diagonal in the \textit{m}-index (azimuthal angle index). \cite[Appendix C]{ref_my} provides an efficient formula that significantly reduces computation time for these cases. However, when $\boldsymbol{d}$ is not parallel to the z-axis, the computations become complex and resource-intensive. To address this, we introduce the rotation matrix of the wavefunctions, which reorients the translation problem from a general direction to alignment along the z-axis, thereby simplifying the computational process.

\begin{figure}[!t]
  \centering
  \includegraphics[]{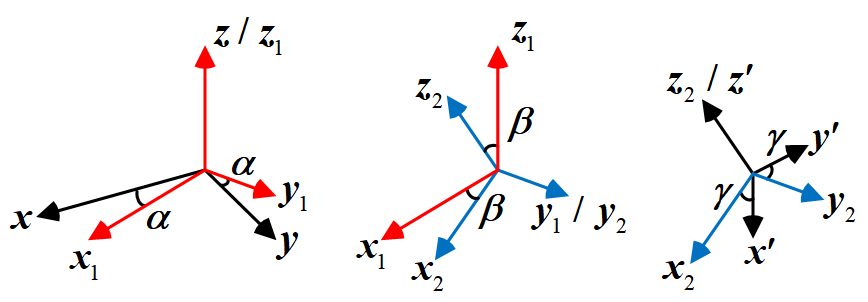}
  \caption{Euler rotation sequence. The coordinate system $(x', y', z')$ is derived by initially rotating the $(x, y, z)$ system about the z-axis by angle $\alpha$, subsequently about the y1-axis by angle $\beta$, and finally about the z2-axis by angle $\gamma$.}
  \label{fEular}
  \vspace{-0.15in}
\end{figure}

Any rotation of the coordinate system can be described using Euler angles $(\alpha,\beta,\gamma)$, where $\alpha$ is an azimuthal rotation about the z-axis, succeeded by a polar rotation $\beta$ about the newly oriented y-axis, and another azimuthal rotation $\gamma$ about the newly oriented z-axis, as illustrated in Fig. \ref{fEular}. We can link the spherical wave expansion vectors $\mathbf{a}/\mathbf{f}$ in the $(x,y,z)$ coordinate system to $\mathbf{a}'/\mathbf{f}'$ in the $(x',y',z')$ coordinate system using a rotation matrix $\bm{\mathcal{D}}=\bm{\mathcal{D}}(\alpha,\beta,\gamma)$ (cf. appendix \ref{app_B} for its definition) as: 
\begin{equation}
  \mathbf{f}'=\bm{\mathcal{D}} \mathbf{f},\mathbf{a}'=\bm{\mathcal{D}} \mathbf{a} \text{ or }
  \mathbf{f}=\bm{\mathcal{D}}^t \mathbf{f}',\mathbf{a}=\bm{\mathcal{D}}^t \mathbf{a}' .
\end{equation}
Note that $\bm{\mathcal{D}}^{-1}=\bm{\mathcal{D}}^t$.

By aligning the z-axis with $\mathbf{d}$, we transform the general translation problem into a z-axis translation problem. The rotation angles $\alpha$ and $\beta$ correspond to the azimuthal and polar angles of the spherical coordinates of $\mathbf{d}$, respectively, and $\gamma = 0$. After performing the z-axis translation, we revert to the original coordinate system through an inverse rotation to resolve the original translation problem. Therefore, we can deduce that
\begin{equation}
  \label{eq12}
  \bm{\mathcal{Y}}\left(k\mathbf{d}\right)=\bm{\mathcal{D}}^t \bm{\mathcal{Y}}^\text{z}\left(kd\right)\bm{\mathcal{D}}
\end{equation}
where $\bm{\mathcal{Y}}^\text{z}\left(kd\right)$ is defined as equation (63) in \cite{ref_my}, referring to translations along the z-axis by distance \textit{d}.

This approach substantially reduces computational time (\textit{e.g.}, computation time is reduced from 30s to 1.2s for a 646-dimensional translation problem, and from 124s to 2.6s for a 1056-dimensional translation problem) because calculating the rotation matrix takes negligible time. Moreover, the rotation matrix is frequency-independent and can be pre-stored, which allows only the translation matrix to be recalculated for different frequencies. For uniformly arranged arrays, where the translation distances remain constant, $\bm{\mathcal{Y}}^\text{z}\left(kd\right)$ needs to be calculated only once. The translation directions are then adjusted by altering the rotation matrix in \eqref{eq12}, further reducing computational burdens.

Additionally, the rotation matrix offers another advantage: obtaining the T-matrix of the structure post-rotation. This concept is analogous to the previous discussion, and the result can be succinctly expressed as
\begin{equation}
  \label{eq13}
  \mathbf{T}^\prime=\bm{\mathcal{D}} ^t\mathbf{T} \bm{\mathcal{D}}
\end{equation}
where $\mathbf{T}^\prime$ is the transition matrix for the rotated structure. \eqref{eq13} provides a new degree of freedom to examine the effects of orientation changes of units within the system, without requiring additional electromagnetic simulations.

\section{Numerical Results: Characteristic Modes for Multi-Structure Systems}
\label{Sec_V}

\subsection{3-cell array}
\label{SecV_A}

\begin{figure}[!t]
  \centering
  \includegraphics[]{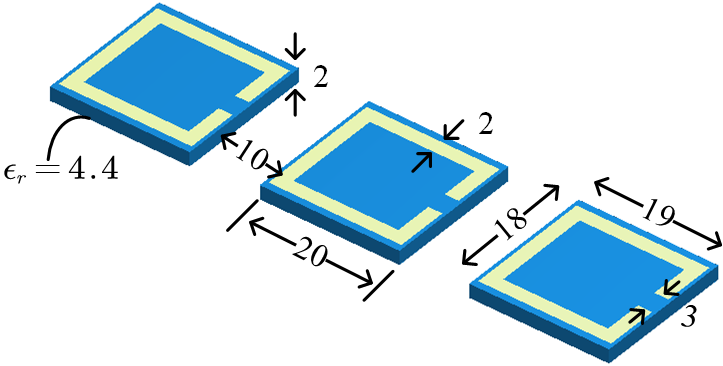}
  \caption{A 3-cell uniform array model with structure dimensions provided in millimeters.}
  \label{f_3array_uniform}
\end{figure}

\begin{figure}[!t]
  \centering
  \includegraphics[]{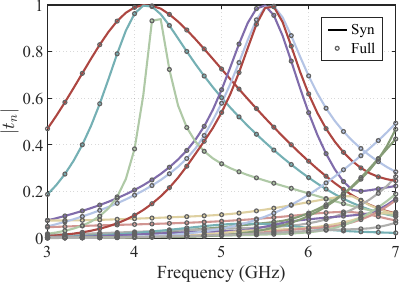}
  \caption{Eigentraces for the 3-cell array radiating in free space. Data labeled ``Syn'' were derived using the synthesis method discussed in Section \ref{Sec_III}; ``Full'' represents results from full-wave simulation.}
  \label{fMS_3array_uniform}
\end{figure}

\begin{figure}[!t]
  \centering
  \subfloat[]{\includegraphics[]{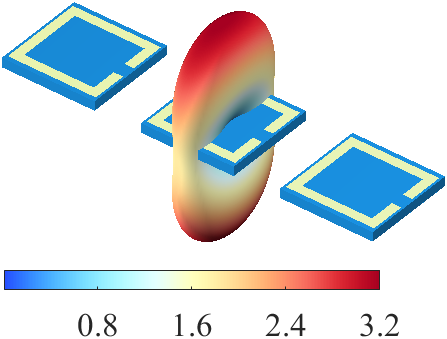}}
  \hfil
  \subfloat[]{\includegraphics[]{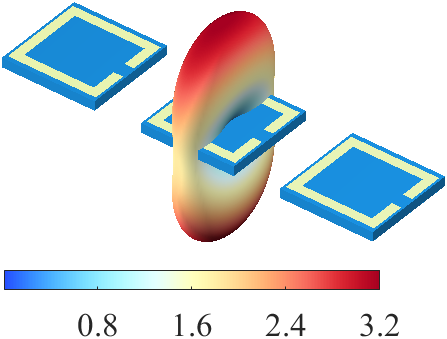}}
  \vfil
  \subfloat[]{\includegraphics[]{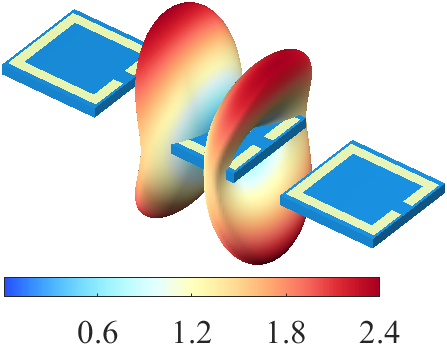}}
  \hfil
  \subfloat[]{\includegraphics[]{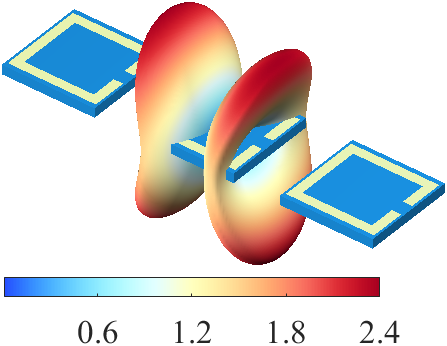}}
  \caption{Radiation patterns at 4.1 GHz. (a) and (b) depict the first mode, while (c) and (d) depict the second mode. The left panels were obtained through full-wave simulation, and the right were through the synthesis method.}
\label{f_array_Rad_all}
\end{figure}

This section highlights the effectiveness of our synthetic approach for calculating characteristic modes through an examination of a uniform 3-cell array. Each cell in the array consists of open-circuit metal rings printed on a substrate, as depicted in Fig. \ref{f_3array_uniform}. Initially, we compute the individual T-matrix of one cell using 646 spherical wavefunctions and 3420 Rao-Wilton-Glisson (RWG) basis functions. Following this, we derive the characteristic modes for the entire array in accordance with the procedure outlined in Sec. \ref{Sec_III}.

Figure \ref{fMS_3array_uniform} presents the eigentraces $|t_n|$ (also referred to as modal significances), where three distinct curves cluster near $|t_n| = 1$, indicating the modal responses of the entire array in free space. Each cell contributes directly to the array's collective behavior. Although the resonant frequencies (where $|t_n| = 1$) of these modes are similar, their radiation characteristics vary markedly. For instance, Fig. \ref{f_array_Rad_all} depicts the radiation patterns of the first two modes at 4.1 GHz: mode 1 exhibits a maximum radiation in the broadside direction, while mode 2 demonstrates a radiation null. The results achieved through our synthetic approach agree with those from full-wave simulations in terms of both eigentraces and radiation patterns. This concurrence validates the accuracy of the method described in Sec. \ref{Sec_III}.

\begin{figure}[!t]
  \centering
  \includegraphics[]{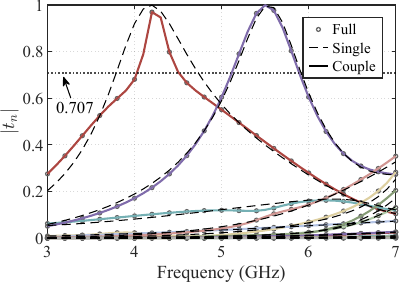}
  \caption{Eigentraces. Lines labeled ``Couple'' account for array environment effects, while ``Single'' considers only the unit response. ``Full'' represents full-wave results. Generally, the 3~dB bandwidth of $|t_n|$ may reveal the practical working bandwidth, marked by a horizontal line at 0.707.}
  \label{fMS_cell}
\end{figure}

\begin{figure}[!t]
  \centering
  \subfloat[]{\includegraphics[]{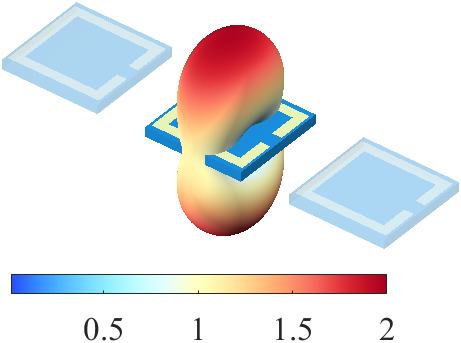}}
  \hfil
  \subfloat[]{\includegraphics[]{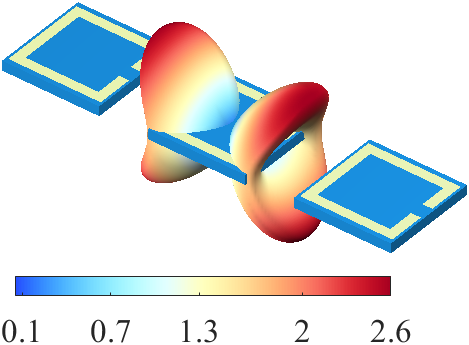}}
  \vfil
  \subfloat[]{\includegraphics[]{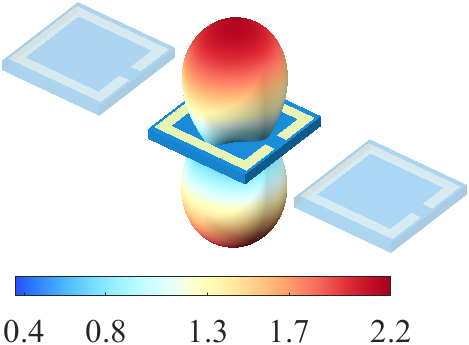}}
  \hfil
  \subfloat[]{\includegraphics[]{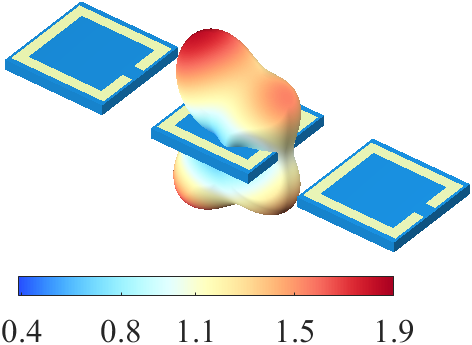}}
  \caption{Radiation patterns. (a), (b) are of mode 1 at 4.2 GHz and (c), (d) are of mode 2 at 5.5 GHz. The left panel is the case when the cell is alone, while the right panel is when in the array environment but with the surrounding cells as background.}
\label{fbenchRad}
\end{figure}

In a subsequent analysis, the setup in Fig. \ref{f_3array_uniform} is modified by designating the cells on both sides as the radiation background for the central cell. The resultant eigentraces, shown as solid lines in Fig. \ref{fMS_cell}, reveal that only two modes resonate within the specified frequency band. The full-wave simulation results (depicted as circles) closely match the solid lines, validating the reliability of our method for ``substructure'' scenarios. When compared to the eigentrace of a single cell in free space (dashed line), the main-mode bandwidth of the array cell is observed to decrease by approximately 13\% due to coupling effects, while the second mode undergoes only slight changes. Additionally, the primary radiation lobes of both modes are no longer directed broadside, instead splitting into different directions, as illustrated in Fig. \ref{fbenchRad}. This finding highlights the importance of accounting for coupling and environmental effects: the array's performance cannot be accurately analyzed or synthesized using the characteristic modes of an isolated cell in free space.

The computational efficiency of the synthetic approach is also noteworthy. The full-wave simulation of the 3-cell array requires simultaneous simulation of all three cells, consuming a total of 83 seconds. In contrast, the synthetic method completes the calculation in just 16 seconds, of which 11 seconds are spent on computing the T-matrix for a single cell. When the number of cells increases to five, the full-wave simulation time exceeds 300 seconds, whereas the synthetic method requires only 24 seconds.

For the moment the array was assumed to have uniform spacing, other methods, like the one detailed in \cite{ref_PCMA}, can utilize shared mesh features to reduce redundant computations. For example, the method in \cite{ref_PCMA} processes a 3-cell array in 33 seconds and a 5-cell array in 52 seconds. While these results are competitive, our synthetic approach still outperforms in efficiency. It is worth noting that we do not announce our method is always superior to \cite{ref_PCMA}. For particularly simple structures, which require fewer RWG basis functions than the number of spherical wavefunctions used in our approach, \cite{ref_PCMA} can indeed be faster.

However, the real advantage of our synthetic method becomes apparent when dealing with arrays that feature non-uniform spacing or elements in rotated orientations. Under such conditions, the efficiency of \cite{ref_PCMA} diminishes markedly as the reusability of impedance matrices decreases, leading to increased computation times. In contrast, our approach necessitates only the computation of a few additional rotation and translation matrices. Since these matrices are analytical, they add small computational overhead. This efficiency makes our method especially suitable for complex array configurations and varying element orientations, where it maintains a consistent performance advantage.

\subsection{Dipole near a sphere}

\begin{figure}[!t]
  \centering
  \includegraphics[]{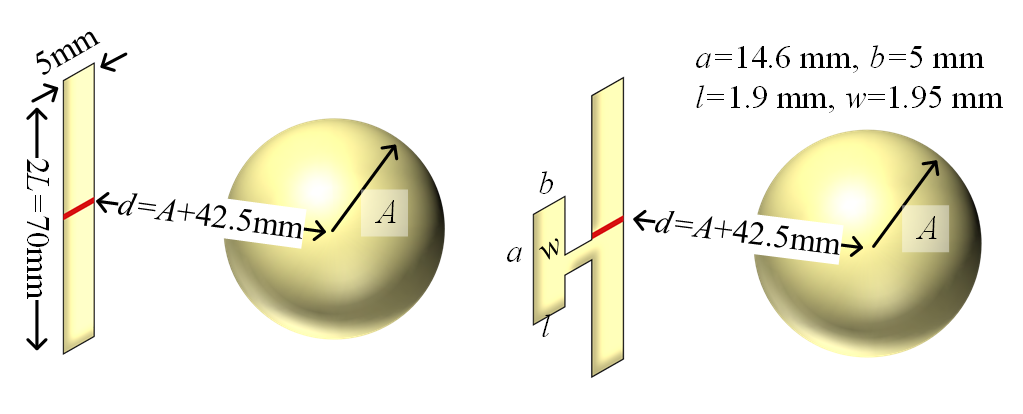}
  \caption{Dipole-sphere model. The left is a regular straight dipole, the right is a modified dipole with T-stub. Both configurations are fed by a discrete port in dipole center.}
  \label{fdipole_sph_model}
\end{figure}

Given that the T-matrix for spherical structures is diagonal and analytically defined {cf. $\S$8 of \cite{ref_scattering_theory}}, the methodology outlined in Sec. \ref{Sec_III} is particularly efficient for analyzing characteristic modes involving spheres, as it eliminates the substantial effort typically required to numerically compute the sphere's T-matrix. We illustrate this with an example analyzing the performance of a dipole in proximity to spherical structures, as shown in the left panel of Fig. \ref{fdipole_sph_model}. Here, the sole requirement is conducting full-wave simulations for the isolated dipole to determine its T-matrix.

\begin{figure}[!t]
  \centering
  \includegraphics[]{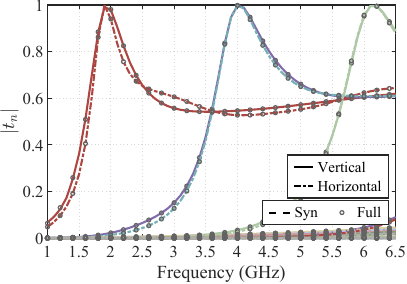}
  \caption{Eigentraces for the regular dipole radiating against the sphere structure. ``Vertical'' indicates the dipole is perpendicular to the sphere, and ``horizontal'' signifies it is parallel to the sphere.}
  \label{fMS_dipole_sph_sub}
\end{figure}

\begin{figure}[!t]
  \centering
  \includegraphics[]{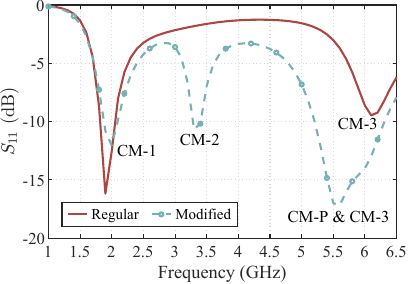}
  \caption{$S_\text{11}$ parameters for the regular and modified dipole when driven by a discrete port centrally.}
  \label{fS11_dipole} 
\end{figure}

\begin{figure}[!t]
  \centering
  \includegraphics[]{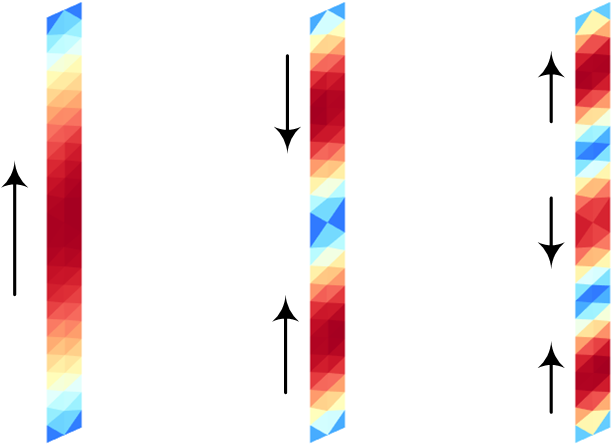}
  \caption{Current distributions on the regular dipole. From left to right is respective the current of CM-1, CM-2 and CM-3.}
  \label{fdipole_current}
\end{figure}

Figure \ref{fMS_dipole_sph_sub} presents the eigentraces for a regular dipole near a perfect electric conductor (PEC) sphere with a diameter of $2A = L$. In the specified frequency range, we identify three resonant modes---labeled ``CM-1'', ``CM-2'' and ``CM-3''---with respective resonant frequencies of 1.9 GHz, 4.1 GHz, and 6.2~GHz. With the dipole aligned vertical to the sphere, the eigentraces alter slightly compared to when it is parallel to the sphere (referred to as the ``horizontal'' configuration). Our synthesized results consistently align with those from full-wave simulations across both configurations. Notably, the T-matrix for the vertically aligned dipole is derivable from the horizontal configurations using \eqref{eq13}. This allows for exploring other dipole polarizations, though these are not pursued further in this paper.

When the dipole is centrally excited, its $S_\text{11}$ parameter, depicted in red in Fig. \ref{fS11_dipole}, shows two resonant bands at 1.9 GHz and 6.2 GHz, corresponding to CM-1 and CM-3, respectively. In line with established principles of characteristic mode theory, this behavior occurs because these modes generate significant current at the dipole's center, which facilitates their excitation, as shown in Fig. \ref{fdipole_current}. In contrast, CM-2 is characterized by a current null at the center, preventing it from being directly excited.

\begin{figure}[!t]
  \centering
  \includegraphics[]{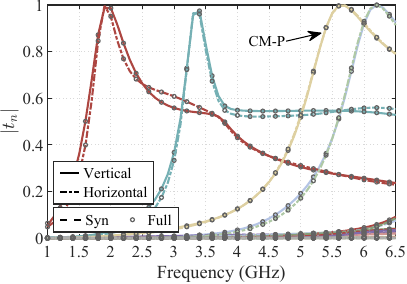}
  \caption{Eigentraces for the modified dipole. ``CM-P'' is a parasitic mode caused by T-stub.}
  \label{fMS_dipoleEx_sph_sub}
\end{figure}

\begin{figure}[!t]
  \centering
  \subfloat[]{\includegraphics[]{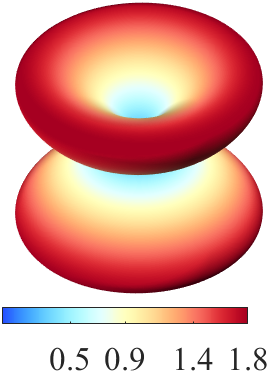}}
  \hfil
  \subfloat[]{\includegraphics[]{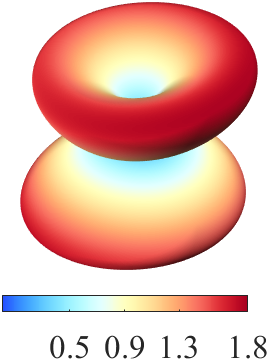}}
  \hfil
  \subfloat[]{\includegraphics[]{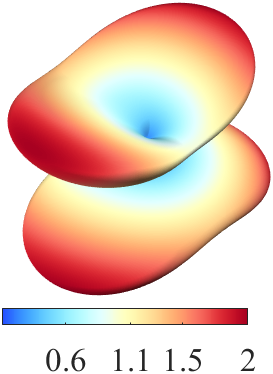}}
  \caption{Radiation patterns of different dipole modes at their respective resonant frequencies: (a) CM-2 of the regular dipole, (b) the second mode of the modified dipole, and (c) the third mode of the modified dipole. To facilitate a clear comparison of these patterns, only the radiation from the dipole currents in free space is presented, omitting any scattering contributions from the spherical background. The resemblance between (a) and (b) suggests that the second mode of the modified dipole likely evolves from the CM-2 of the regular dipole.}
\label{fCM_Rad_dipole}
\end{figure}

To address this, we introduced a small T-stub at the center of the dipole, depicted in the right panel of Fig. \ref{fdipole_sph_model}. This modification aims to shift the current null of CM-2 from the dipole's center to the stub. The updated eigentraces, presented in Fig. \ref{fMS_dipoleEx_sph_sub}, now exhibit four resonant modes. It is apparent that the first and last curves correspond to CM-1 and CM-3 in Fig. \ref{fMS_dipole_sph_sub}, respectively, as they remain largely unchanged. The radiation patterns for the second and third traces are depicted in Fig. \ref{fCM_Rad_dipole}b and \ref{fCM_Rad_dipole}c, respectively. Upon comparison with Fig. \ref{fCM_Rad_dipole}a, it can be inferred that the T-stub reduces the resonant frequency of the original CM-2 to 3.3 GHz. Consequently, the new mode, CM-P, which resonates at 5.7 GHz, is identified as a parasitic mode induced by the T-stub. Given the T-stub's smaller size compared to the dipole, CM-P exhibits a higher resonant frequency.

\begin{figure}[!t]
  \centering
  \includegraphics[]{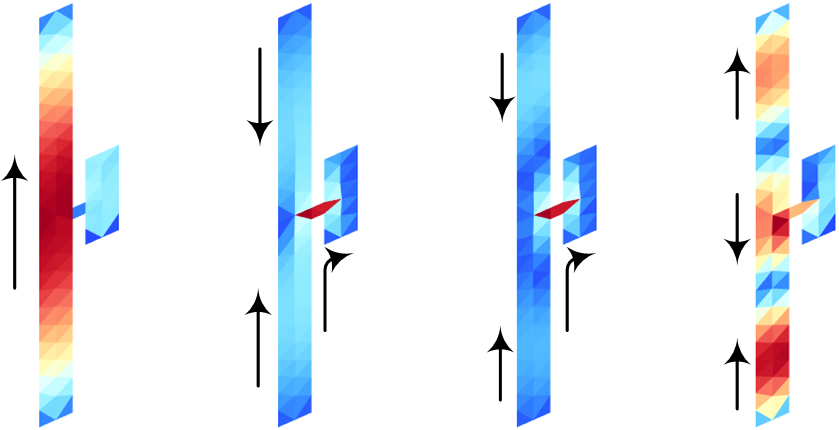}
  \caption{Current distributions on the modified dipole. From left to right is respective the current of CM-1, CM-2, CM-P, and CM-3.}
  \label{fdipoleEx_current}
\end{figure}

\begin{figure}[!t]
  \centering
  \subfloat[]{\includegraphics[]{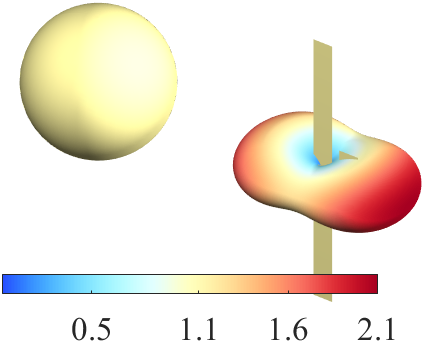}}
  \hfil
  \subfloat[]{\includegraphics[]{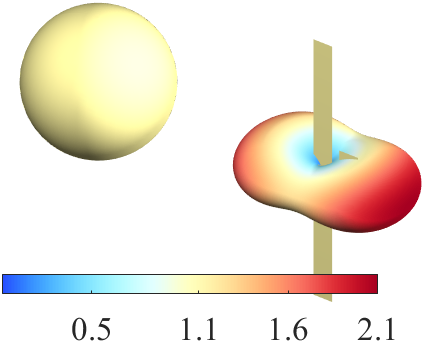}}
  \vfil
  \subfloat[]{\includegraphics[]{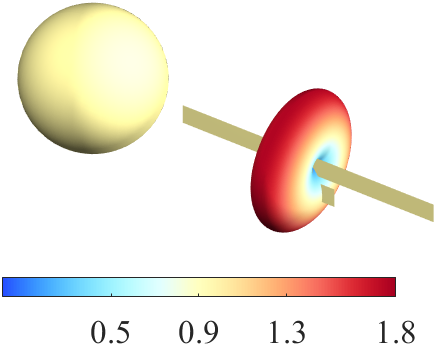}}
  \hfil
  \subfloat[]{\includegraphics[]{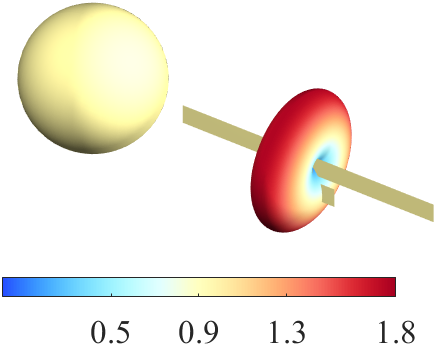}} 
  \caption{Radiation pattern of the main mode at 1.9 GHz. (a) and (b) depict the horizontal case, while (c) and (d) show the vertical case. The left panel in each pair is obtained through full-wave simulation, and the right panel through synthesis method. In this figure, the sphere diameter is $2A=L$.}
\label{fdipole_sph_Rad}
\end{figure}

\begin{figure}[!t]
  \centering
  \subfloat[]{\includegraphics[]{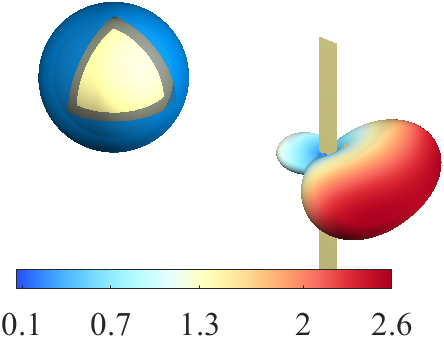}}
  \hfil
  \subfloat[]{\includegraphics[]{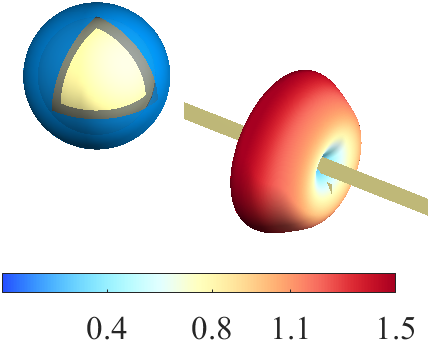}}
  \vfil
  \subfloat[]{\includegraphics[]{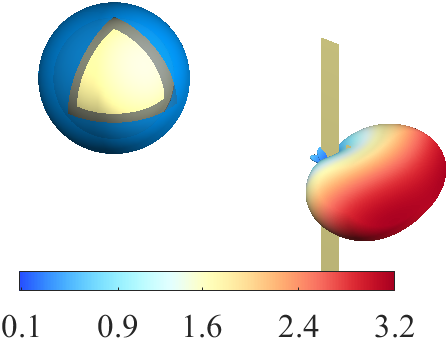}}
  \hfil
  \subfloat[]{\includegraphics[]{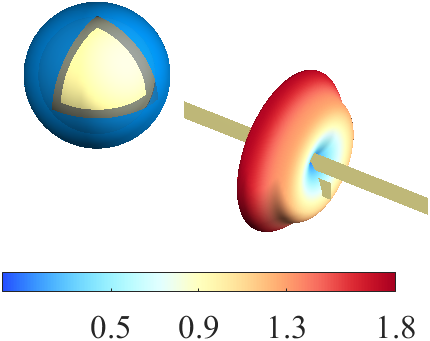}}
  \vfil
  \subfloat[]{\includegraphics[]{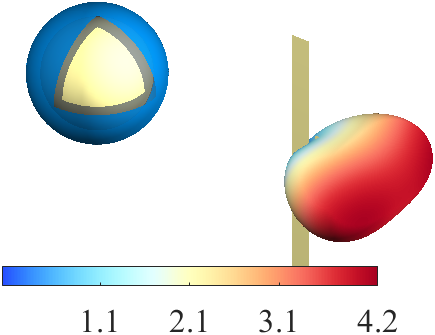}}
  \hfil
  \subfloat[]{\includegraphics[]{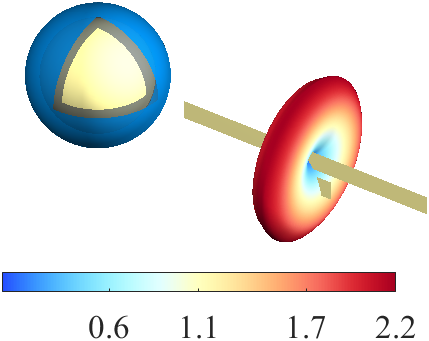}}
  \caption{Radiation patterns. (a), (c), and (e) represent the horizontal case, while (b), (d), and (f) represent the vertical case. From the top to the bottom, the diameter of the layered sphere increases progressively from $2L$ to $5L$, and finally to $20L$. To facilitate clear observation of the orientation of the radiation lobes, \textit{we scale the images of the spheres to the same apparent size in all figures}.}
\label{fdipole_sph_inc}
\end{figure}

Figure \ref{fdipoleEx_current} depicts the current distribution for each mode of the modified dipole. As none of the modes exhibit a current null at the dipole's center, they are well-suited for center-feed excitation. The resulting $S_{\text{11}}$ for the modified dipole, displayed in green in Fig. \ref{fS11_dipole}, highlights the advantages of this modification by introducing an additional operating band and significantly expanding the high-frequency bandwidth through the integration of CM-P with CM-3. Furthermore, Fig. \ref{fdipole_sph_Rad} presents the main-mode radiation patterns for the dipole near a PEC sphere in both horizontal and vertical orientations, aligning closely with results from center-driven simulations and thereby further validating our design modifications. Analytical results from the T-matrix are available for spheres composed of various materials---such as dielectric, layered, anisotropic, or biisotropic---enabling discussions on different spherical backgrounds. For instance, Fig. \ref{fdipole_sph_inc} depicts the radiation pattern of CM-1 from the modified dipole adjacent to a layered spherical structure (comprising three layers with radii of $A$, $0.8A$, and $0.64A$ and relative permittivities of 38 and 15 for the two outer layers, respectively, with the innermost layer being PEC). As the sphere's diameter increases, the directivity of the horizontally oriented dipole improves, and backward radiation decreases, which corresponds with our engineering intuition. In contrast, for the vertically oriented dipole, the maximum radiation initially leans towards the sphere and then progressively shifts away. Notably, in the bottom panel of Fig. \ref{fdipole_sph_inc}, where the sphere's diameter is 20 times the length of the dipole, this may inform studies of ground effects for structures near the Earth, which may be approximated as a layered sphere.

\section{Conclusion and Discussion}

This paper introduces a method to accelerate the decomposition of characteristic modes for structural combinations by utilizing T-matrices of individual structures. Our numerical results validate the method's accuracy and demonstrate its substantial computational advantages through the example of a 3-cell array. Additionally, the analytical nature of the T-matrix for spheres allows for straightforward exploration of the variations in characteristic modes resulting from changes in the position and orientation of structures near spherical bodies, such as the Earth.

While the T-matrix effectively relates the incident and scattering fields in the external region of any sphere circumscribing the structure, it does not account for the internal regions. Therefore, this study posits that different structures can be encapsulated within distinct, non-intersecting spheres. This realization underscores that further research is needed to develop methods for accelerating characteristic mode decomposition in tightly coupled arrays or structures in contact with one another.

\begin{appendices}
\section{Schur complement method}
\label{app_A}
The Schur complement method is a powerful technique for solving matrix equations like
\begin{equation}
  \label{eq_a1}
  \mathbf{AX}=\mathbf{B}
\end{equation}
by utilizing the sub-blocks of the matrices involved.

Consider bifurcating \eqref{eq_a1} as
\begin{equation}
  \left[ \begin{matrix}
    \mathbf{A}_{11}&		\mathbf{A}_{12}\\
    \mathbf{A}_{21}&		\mathbf{A}_{22}\\
  \end{matrix} \right] \begin{bmatrix}
    \mathbf{X}_1\\
    \mathbf{X}_2\\
  \end{bmatrix} =\begin{bmatrix}
    \mathbf{B}_1\\
    \mathbf{B}_2\\
  \end{bmatrix} 
\end{equation}
which gives the solution for $\mathbf{X}_2=\mathbf{A}_{22}^{-1}\left\{ \mathbf{B}_2-\mathbf{A}_{21}\mathbf{X}_1 \right\} $. This leads to a reduced equation for $\mathbf{X}_1$:
\begin{equation}
  \tilde{\mathbf{A}}\mathbf{X}_1=\tilde{\mathbf{B}}
\end{equation}
where $\tilde{\mathbf{A}}=\mathbf{A}_{11}-\mathbf{A}_{12}\mathbf{A}_{22}^{-1}\mathbf{A}_{21}$, $\tilde{\mathbf{B}}=\mathbf{B}_1-\mathbf{A}_{12}\mathbf{A}_{22}^{-1}\mathbf{B}_2$. Collapsing these results, we have
\begin{equation}
  \label{eq_a4}
  \mathbf{X}=\mathbf{A}^{-1}\mathbf{B}=\begin{bmatrix}
    \mathbf{0}\\
    \mathbf{A}_{22}^{-1}\mathbf{B}_2\\
  \end{bmatrix} +\begin{bmatrix}
    \mathbf{1}\\
    -\mathbf{A}_{22}^{-1}\mathbf{A}_{21}\\
  \end{bmatrix} \tilde{\mathbf{A}}^{-1}\tilde{\mathbf{B}}.
\end{equation}

In the context of this paper, our objective is to determine the term $\left( \mathbf{1}-\tilde{\mathbf{T}}\tilde{\bm{\mathcal{Y}}} \right) ^{-1} \tilde{\bm{\mathcal{R}}}^t$ in \eqref{eq10}, where $\left( \mathbf{1}-\tilde{\mathbf{T}}\tilde{\bm{\mathcal{Y}}} \right)$ serves as $\mathbf{A}$ in \eqref{eq_a4}, and $\mathbf{A}_{22}$ contains matrices solely associated with background structures. In the process of calculating the background structures' T-matrix, the term $\mathbf{A}_{22}^{-1}\mathbf{B}_2$ has been previously determined, necessitating only additional effort to find $\tilde{\mathbf{A}}^{-1}$. Note that the scale of $\tilde{\mathbf{A}}^{-1}$ is relatively small, limited by the remaining structures after excluding the background structures.

\section{Rotation Matrix}
\label{app_B}
There are many ways \cite{ref_rotation1,ref_rotation2} to express the element of the rotation matrix $\bm{\mathcal{D}}\left(\alpha, \beta, \gamma \right)$, the most concise is in the form of matrix products:
\begin{equation}
  \label{eq_a5}
  \begin{split}
    \bm{\mathcal{D}} _{nn'} =\delta _{\tau \tau '}&\delta _{ll'}\sqrt{\frac{\varepsilon _m\varepsilon _{m'}}{4}}\left( -1 \right) ^{m+m'}
\\
&\times \left[ \begin{matrix}
	\cos \left( m\gamma \right)&		\sin \left( m\gamma \right)\\
	-\sin \left( m\gamma \right)&		\cos \left( m\gamma \right)\\
\end{matrix} \right] 
\\
&\times \left[ \begin{matrix}
	A_{mm'}^{l}\left( \beta \right)&		\\
	&		B_{mm'}^{l}\left( \beta \right)\\
\end{matrix} \right] 
\\
&\times \left[ \begin{matrix}
	\cos \left( m'\alpha \right)&		\sin \left( m'\alpha \right)\\
	-\sin \left( m'\alpha \right)&		\cos \left( m'\alpha \right)\\
\end{matrix} \right] 
\end{split}
\end{equation}
where the subscript $n=\tau\sigma l m$ is a composite index \cite{ref_myGSM,ref_scattering_theory}. The products of the three matrices in \eqref{eq_a5} result in a $2\times2$ matrix with the row index as $\sigma$ and the column index as $\sigma'$. $\bm{\mathcal{D}} _{nn'}$ should selected a value from this $2\times2$ matrix corresponding to the index $\sigma\sigma'$ (\textit{i.e.}, ee, eo, oe and oo). The $\delta$ in \eqref{eq_a5} represents kronecker delta, and $\varepsilon_m=2-\delta_{m0}$. The coefficients
\begin{equation}
  \begin{split}
    &A_{mm'}^{l}\left( \beta \right) =d_{mm'}^{l}\left( \beta \right) +\left( -1 \right) ^{m'}d_{m-m'}^{l}\left( \beta \right) \\
    &B_{mm'}^{l}\left( \beta \right) =d_{mm'}^{l}\left( \beta \right) -\left( -1 \right) ^{m'}d_{m-m'}^{l}\left( \beta \right) 
  \end{split}
\end{equation}
and the term $d_{mm'}^{l}$ are given by
\begin{equation}
  \begin{split}
    d_{mm'}^{l}\left( \beta \right) &=\sqrt{\frac{\left( l+m \right) !\left( l-m \right) !}{\left( l+m' \right) !\left( l-m' \right) !}}\\
    &\times \cos ^{m+m'}\left( \frac{\beta}{2} \right) \sin ^{m-m'}\left( \frac{\beta}{2} \right)\\
    &\times P_{l-m}^{\left( m-m',m+m' \right)}\left( \cos \beta \right) 
  \end{split}
\end{equation}
where $P_{n}^{\left( a,b \right)}(x)$ denotes the Jacobi polynomials, typically computed using a three-term recurrence relation. Some useful properties include:
\begin{equation}
  \begin{split}
    &d_{mm'}^{l}(0)=\delta_{mm'}\\
    &d_{mm'}^{l}(\pi)=(-1)^{l+m'}\delta_{m-m'}\\   
    &\bm{\mathcal{D}}^{-1}\left(\alpha, \beta, \gamma \right)=\bm{\mathcal{D}}\left(-\gamma, -\beta, -\alpha \right)=\bm{\mathcal{D}}^t\left(\alpha, \beta, \gamma \right)\\
    &\bm{\mathcal{D}}\left(\alpha, \beta, \gamma \right)=\bm{\mathcal{D}}\left(\gamma \right)\bm{\mathcal{D}}\left(\beta \right)\bm{\mathcal{D}}\left(\alpha \right).
  \end{split}
\end{equation}
\end{appendices}

\begin{IEEEbiography}[{\includegraphics[width=1in,height=1.25in,clip,keepaspectratio]{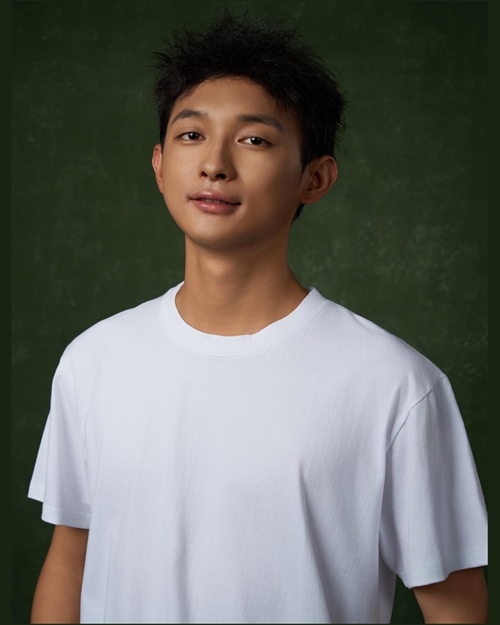}}]{Chenbo Shi}
  Chenbo Shi was born in 2000 in China. He received his Bachelor's degree from the University of Electronic Science and Technology of China (UESTC) in 2022. He is currently pursuing his Ph.D. at the same institution. His research interests include electromagnetic theory, characteristic mode theory, and computational electromagnetics. 
  
  Chenbo has been actively involved in several research projects and has contributed to publications in these areas. His work aims to advance the understanding and application of electromagnetic phenomena in various technological fields.
  \end{IEEEbiography}
    
  \begin{IEEEbiography}[{\includegraphics[width=1in,height=1.25in,clip,keepaspectratio]{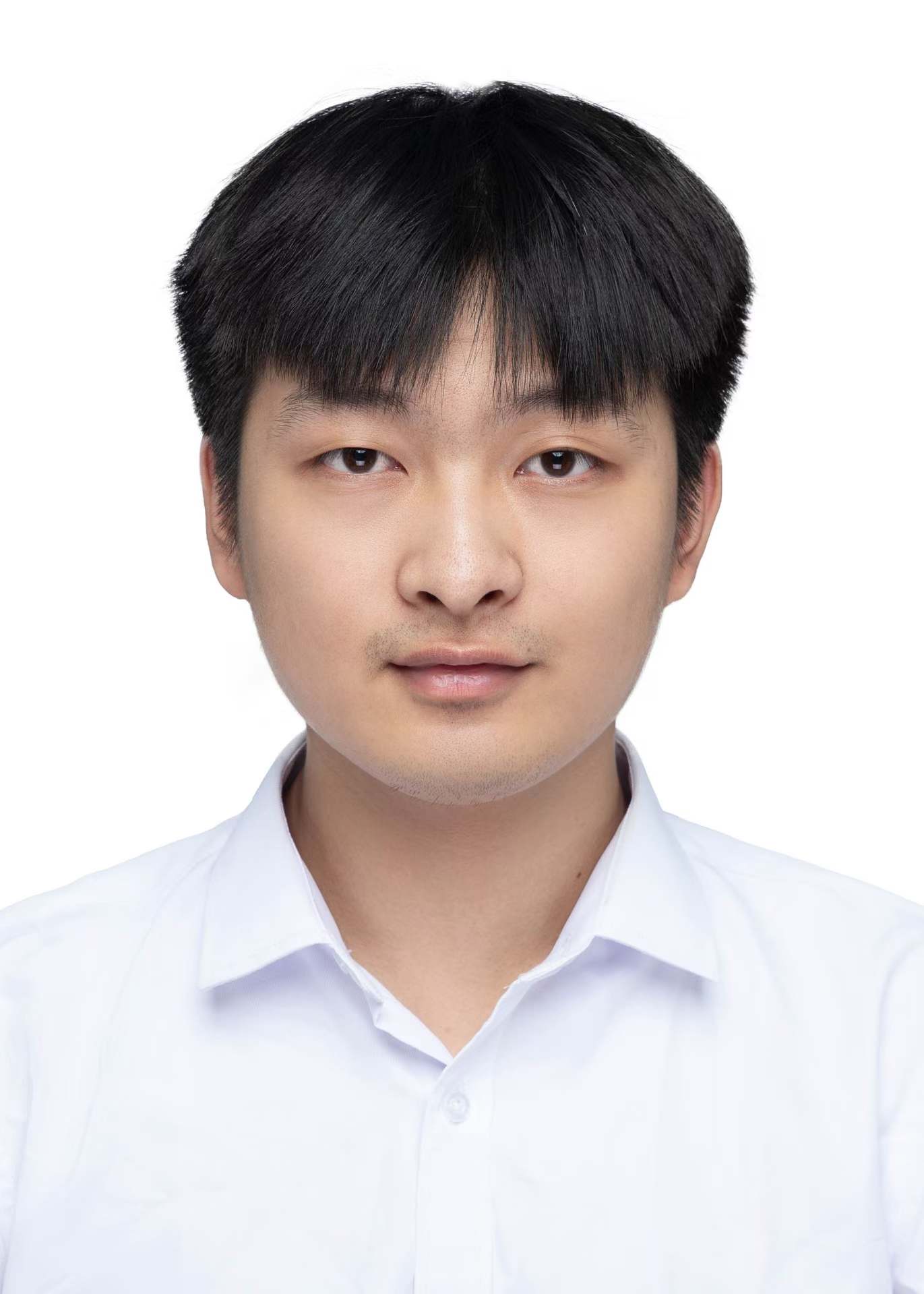}}]{Xin Gu}
  received the B.E. degree from Chongqing University of Posts and Telecommunications (CQUPT), ChongQing, China, in 2022. He is currently pursuing the M.S. degree with the School of Electronic Science and Engineering, University of Electronic Science and Technology of China (UESTC), Chengdu, China.
    
  His research interests include electromagnetic theory and electromagnetic measurement techniques
  \end{IEEEbiography}
  
  \begin{IEEEbiography}[{\includegraphics[width=1in,height=1.25in,clip,keepaspectratio]{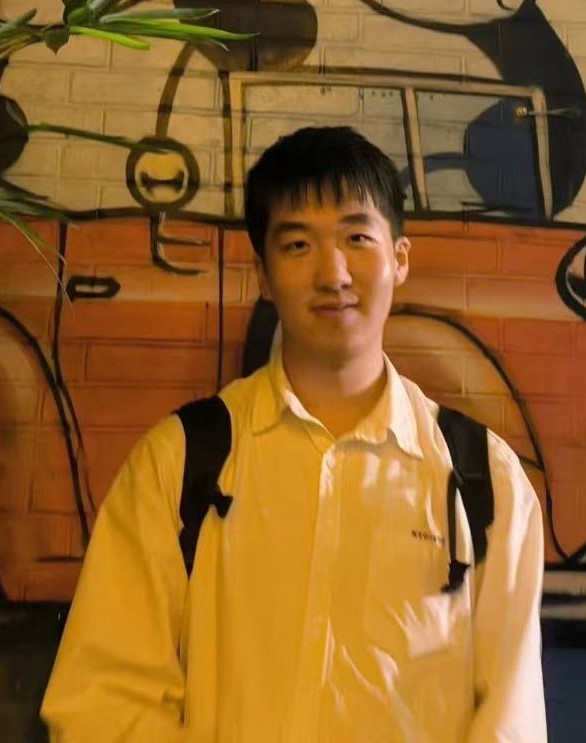}}]{Shichen Liang}
  received the B.E. degree from Beijing University of Chemical Technology (BUCT), Beijing, China, in 2022. He is currently pursuing the M.S. degree with the School of Electronic Science and Engineering, University of Electronic Science and Technology of China (UESTC), Chengdu, China. 
  
  His research interests include electromagnetic theory and electromagnetic measurement techniques.
  \end{IEEEbiography}

  \begin{IEEEbiography}[{\includegraphics[width=1in,height=1.25in,clip,keepaspectratio]{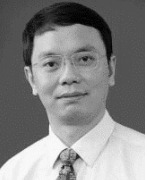}}]{Jin Pan}
    received the B.S. degree in electronics and communication engineering from the Radio Engineering Department, Sichuan University, Chengdu, China, in 1983, and the M.S. and Ph.D. degrees in electromagnetic field and microwave technique from the University of Electronic Science and Technology of China (UESTC), Chengdu, in 1983 and 1986, respectively. 
    
    From 2000 to 2001, he was a Visiting Scholar in electronics and communication engineering with the Radio Engineering Department, City University of Hong Kong. He is currently a Full Professor with the School of Electronic Engineering, UESTC. 
    
    His current research interests include electromagnetic theories and computations, antenna theories, and techniques, field and wave in inhomogeneous media, and microwave remote sensing theories and its applications. 
    \end{IEEEbiography}

  \begin{IEEEbiography}[{\includegraphics[width=1in,height=1.25in,clip,keepaspectratio]{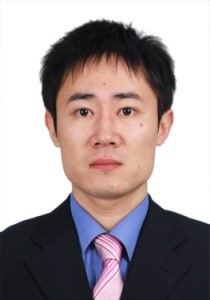}}]{Le Zuo}
  received the B.Eng., M.Eng. and Ph.D. degrees in electromagnetic field and microwave techniques from the University of Electronic Science and Technology of China (UESTC), in 2004, 2007 and 2018, respectively. 
  
  From 2017 to 2018, he was a Research Associate with the School of Electrical and Electronic Engineering, Nanyang Technological University, Singapore. He is currently a Research Fellow with the 29th Institute of the China Electronics Technology Group, Chengdu, China. His research interests include antenna theory and applications.
  \end{IEEEbiography}

\end{document}